\documentclass[aps,prb,twocolumn,a4paper,floatfix,superscriptaddress,showpacs,showkeys]{revtex4-1}

\usepackage[usenames,dvipsnames]{color}

\usepackage{dcolumn,amsmath,xspace}

\usepackage{graphicx,color}
 \usepackage{multirow}

\usepackage{epstopdf}

\newcommand{\sixS}{\ensuremath{(6\!\times\!6)_\text{SiC} ~}}
\newcommand{\six}{\ensuremath{(6\!\times\!6) ~}}
\newcommand{\degC}{\ensuremath{^{\circ}\text{C }}}
\newcommand{\sixrt}{\ensuremath{(6\sqrt{3}\!\times\!6\sqrt{3})_\text{SiC}\text{R}30^\circ}~}

\newcommand{\CB}{C\textsubscript{B}}
\newcommand{\BGo}{BG\textsubscript{o}}
\newcommand{\BGML}{BG\textsubscript{ML}}
\newcommand{\Stwo}{S\textsubscript{2}}
\newcommand{\Sone}{S\textsubscript{1}}
\newcommand{\SG}{S\textsubscript{G}}
\newcommand{\CBp}{C'\textsubscript{B}}
\newcommand{\SML}{S\textsubscript{ML}}

\begin{document}
\title{The structure and evolution of semiconducting buffer graphene grown on SiC(0001)}

\author{M.~Conrad}
\affiliation{The Georgia Institute of Technology, Atlanta, Georgia 30332-0430, USA}
\author{J. Rault}
\author{Y. Utsumi}
\affiliation{Synchrotron SOLEIL, L'Orme des Merisiers, Saint-Aubin, 91192 Gif sur Yvette, France}
\author{Y. Garreau}
\affiliation{Synchrotron SOLEIL, L'Orme des Merisiers, Saint-Aubin, 91192 Gif sur Yvette, France}
\affiliation{Universit\'{e} Paris Diderot, Sorbonne-Paris-Cit\'{e}, MPQ, UMR 7162 CNRS, B\^{a}timent Condorcet, Case 7021, 75205 Paris Cedex 13, France}
\author{A. Vlad} 
\author{A. Coati}
\author{J.-P. Rueff}
\affiliation{Synchrotron SOLEIL, L'Orme des Merisiers, Saint-Aubin, 91192 Gif sur Yvette, France}
\author{P.F.~Miceli}
\affiliation{Department of Physics and Astronomy, University of Missouri-Columbia, Columbia, MO 65211}
\author{E.H.~Conrad}\email[email: ]{edward.conrad@physics.gatech.edu}
\affiliation{The Georgia Institute of Technology, Atlanta, Georgia 30332-0430, USA}

\begin{abstract}
Using highly controlled coverages of graphene on SiC(0001), we have studied the structure of the first graphene layer that grows on the SiC interface.  This layer, known as the buffer layer, is semiconducting. Using x-ray reflectivity and x-ray standing waves analysis we have performed a comparative study of the buffer layer structure with and without an additional monolayer graphene  layer above it.  We show that no more than 26\% of the buffer carbon is covalently bonded to Si in the SiC interface.  We also show that the top SiC bilayer is Si depleted and is the likely the cause of the incommensuration previously observed in this system. When a monolayer graphene layer forms above the buffer, the buffer layer becomes less corrugated with signs of a change in the bonding geometry with the SiC interface.  At the same time, the entire SiC interface becomes more disordered, presumably due to entropy associated with the higher growth temperature.  
\end{abstract}
\vspace*{4ex}

\maketitle
\newpage

\section{Introduction}
The first graphene ``buffer" layer that grows on the SiC(0001) surface is one of the most important examples of functionalized graphene. It is normally in a semiconducting state due to self functionalization caused by sp\textsuperscript{3} bonding to silicon atoms in the SiC interface.\cite{Emtsev_PRB_08,Nevius_PRL_15}  The buffer's bandgap can be increased by additional functionalization with fluorine\cite{Walter_PRB_16} or transformed to a metallic graphene form by H\textsubscript{2} intercalation that breaks the sp\textsuperscript{3} bonding to the interface Si.\cite{Riedl_PRL_09}  In fact numerous studies have shown that the buffer graphene's electronic properties can be altered by changing the interfacial Si bonds,\cite{Mattausch_PRL_07,Varchon_PRL_07,Kim_PRL_08,Varchon_PRB_2008,Riedl_PRL_09,Nevius_PRL_15,Walter_PRB_16,Conrad_NL_17} implying that the buffer's electronic properties can in principle be modified in a controlled fashion.  However, the level of understanding necessary to systematically alter the buffer's properties has remained illusive because structural details, like the number of C-Si bonds and their geometry, are simply not well understood. 

The bonding geometry problem is underscored by the number of different states calculations predict for the  buffer.  Ab initio calculations using a $(\sqrt{3}\!\times\!\sqrt{3})_\text{SiC}$R30 cell find a wide bandgap buffer while calculations on larger, experimentally observed \sixrt cells\cite{VanBommel_SS_75,Forbeaux_PRB_98} find metallic states running through the Fermi Energy  ($E_F$).\cite{Mattausch_PRL_07,Varchon_PRL_07,Kim_PRL_08} The applicability of these early calculations is problematic because they all assumed that the SiC surface is bulk terminated,\cite{Kim_PRL_08,Varchon_PRB_2008,Sforzini_PRL_15} an assumption that we now know is incorrect.  Recent x-ray diffraction studies have demonstrated that the buffer-SiC interface is not commensurate with SiC.\cite{Conrad_NL_17}  Instead, the system has an incommensurate (IC) modulation period $\lambda\!=\!6(1+\delta)a_\text{SiC}$ ($\delta\!=\!0.037$) that is close to the \sixS subcell of the the \sixrt cell.
Tight binding (TB) calculations, using an IC in-plane distortion of a bulk terminated surface, show that the distortion can open a bandgap similar to that measured by angle resolved photoemission (ARPES).\cite{Conrad_NL_17,Nevius_PRL_15}  
However, before more sophisticated ab initio calculations on the IC structure can be attempted, some details of the interface structure will be needed to limit the parameter space for these taxing computations.

Experimentally determining the buffer's structure is also problematic. The first experimental structural studies were done on samples grown in ultra high vacuum (UHV).\cite{Hass_JPCM_08} These samples suffered from both reduced long range order\cite{Charrier_JApPhys_02,Hass_APL_06,Gao_PRB_08} and poor control of both the number of graphene layers and their lateral distribution.\cite{Charrier_JApPhys_02,Riedl_PRB_07} Sample uniformity turns out to be extremely important because both the buffer's electronic and structural properties are now known to change when monolayer (ML) graphene grows above the buffer.\cite{Conrad_NL_17} This means that measured structures on nonuniform films represent some unknown average of two different structures.

\begin{figure}
\includegraphics[angle=0,width=7.5cm,clip]{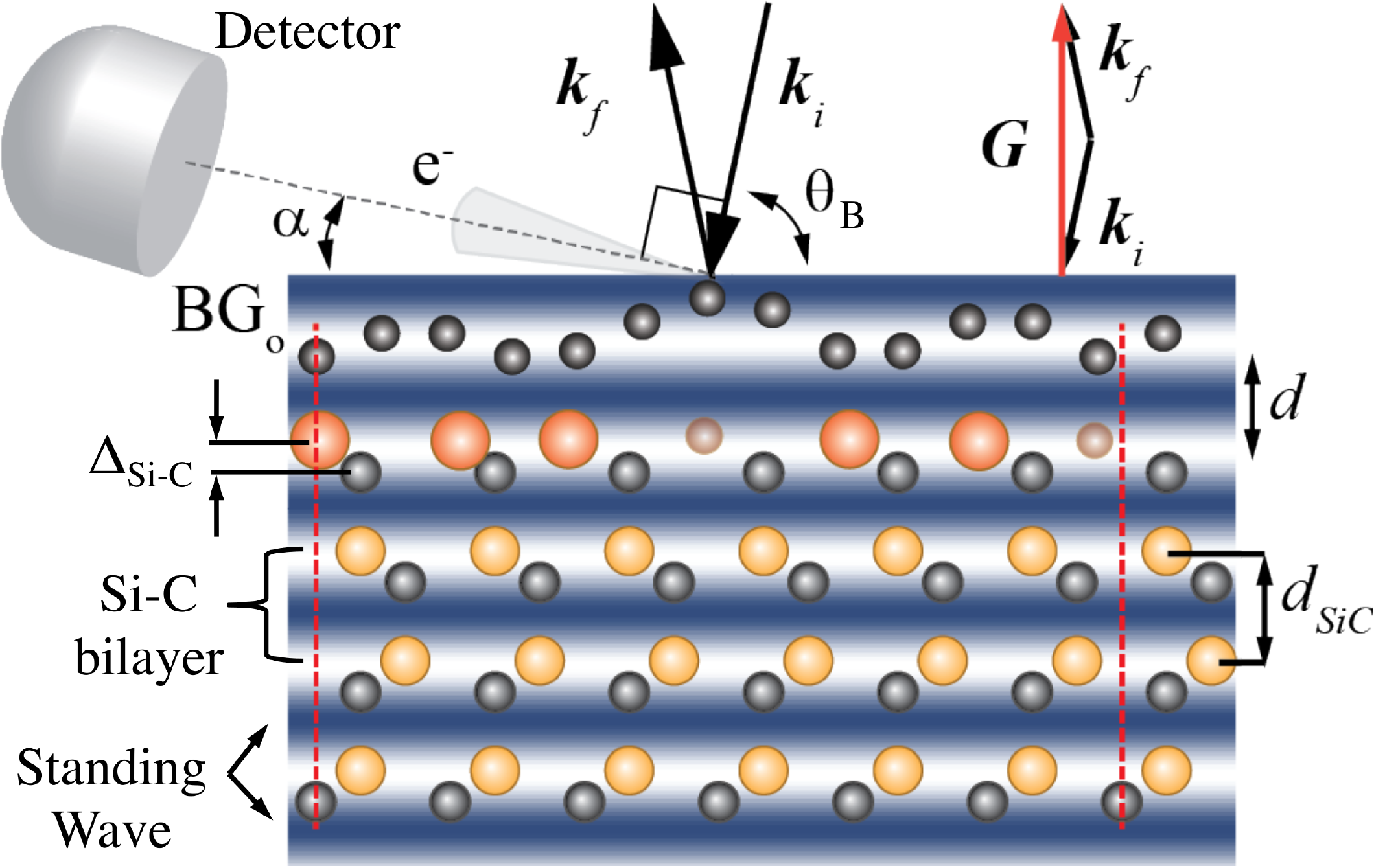}
\caption{Schematic of the XSW geometry on the SiC(0001) surface.  Grey circles show the buckled \BGo~buffer layer above the top SiC bilayer (Yellow and grey circles are Si and C, respectively, in the SiC lattice). The photoelectrons are measured by the detector at a takeoff angle of $\alpha$.  The standing wave period, $d$, is shown using the $G=2\pi/d_\text{SiC}$ Bragg reflection.}
\label{F:XSW_SCH}
\end{figure}

In this work, we use x-ray standing wave (XSW) and x-ray reflectivity (XRR) measurements to study two types of buffer graphene films grown on SiC(0001): a buffer-only film (\BGo) and a buffer graphene film (\BGML) that has a monolayer graphene layer grown above it.  Because of the improved thickness control and the layer uniformity achievable in silicon sublimation controlled RF furnace grown graphene, we are able to discriminate structural changes in these two different types of buffer graphene. We show that the buffer structure and the bonding to the SiC are very different with and without a monolayer graphene layer grown above the buffer layer.  While these differences help explain recent in-plane x-ray diffraction results,\cite{Conrad_NL_17} they complicated previous XSW analysis that used multilayer films and led to a misidentification of the buffer-SiC bonding component in the buffer's C1s spectrum.\cite{Emery_SW_13}  We also show that the buffer-only film has a large vertical corrugation and that its close distance to the substrate indicates a strong sp\textsuperscript{3} C-Si bond.  The buffer's C-Si bond length increases and the corrugation amplitude becomes smaller when the ML grows above the buffer, indicating a change in the distribution of graphene-Si bonds to the SiC interface. Finally, we confirm that the Si concentration in the last Si-C bilayer [see Fig.~\ref{F:XSW_SCH}] is reduced as previously reported.\cite{Emery_SW_13} Rather than being a growth artifact as perviously conjectured, we show that Si vacancy concentration is an equilibrium structure of the top SiC bilayer.  We find 25\% SiC vacancies in the top SiC bilayer (compared to the bulk value) for both \BGo~and \BGML~films.  This result helps put an upper limit of 26\% on the number of buffer-carbon atoms bonded to silicon at the interface. The reduced Si concentration is coupled with a vertical compression in the Si-C bilayer below the buffer, suggesting that the Si vacancy concentration may help drive the incommensurate structure of the \BGo~and SiC interface. 

\section{Experimental Methods}
\subsection{Sample Preparation}
The substrates used in these studies were n-doped CMP polished on-axis 4H-SiC(0001). The graphene was grown in a confinement controlled silicon sublimation furnace (CCS).\cite{WaltPNAS} In the CCS method, graphene growth is a function of temperature, time, and crucible geometry that sets the silicon vapor pressure. With the current crucible design,\cite{Nevius_phd} a single semiconducting \BGo~graphene layer grows at a temperature of 1400\degC after 30~min. Heating at 1560\degC for 20~min causes the \BGo~layer to transform into a graphene ML as a new \BGML~buffer layer forms below the ML. Prior to XSW and XRR experiments,
the \BGo~and \BGML+ML samples were characterized by x-ray photoemission spectroscopy (XPS) and Raman to determine film quality and ML coverage. 

\begin{figure}
\includegraphics[angle=0,width=7.0cm,clip]{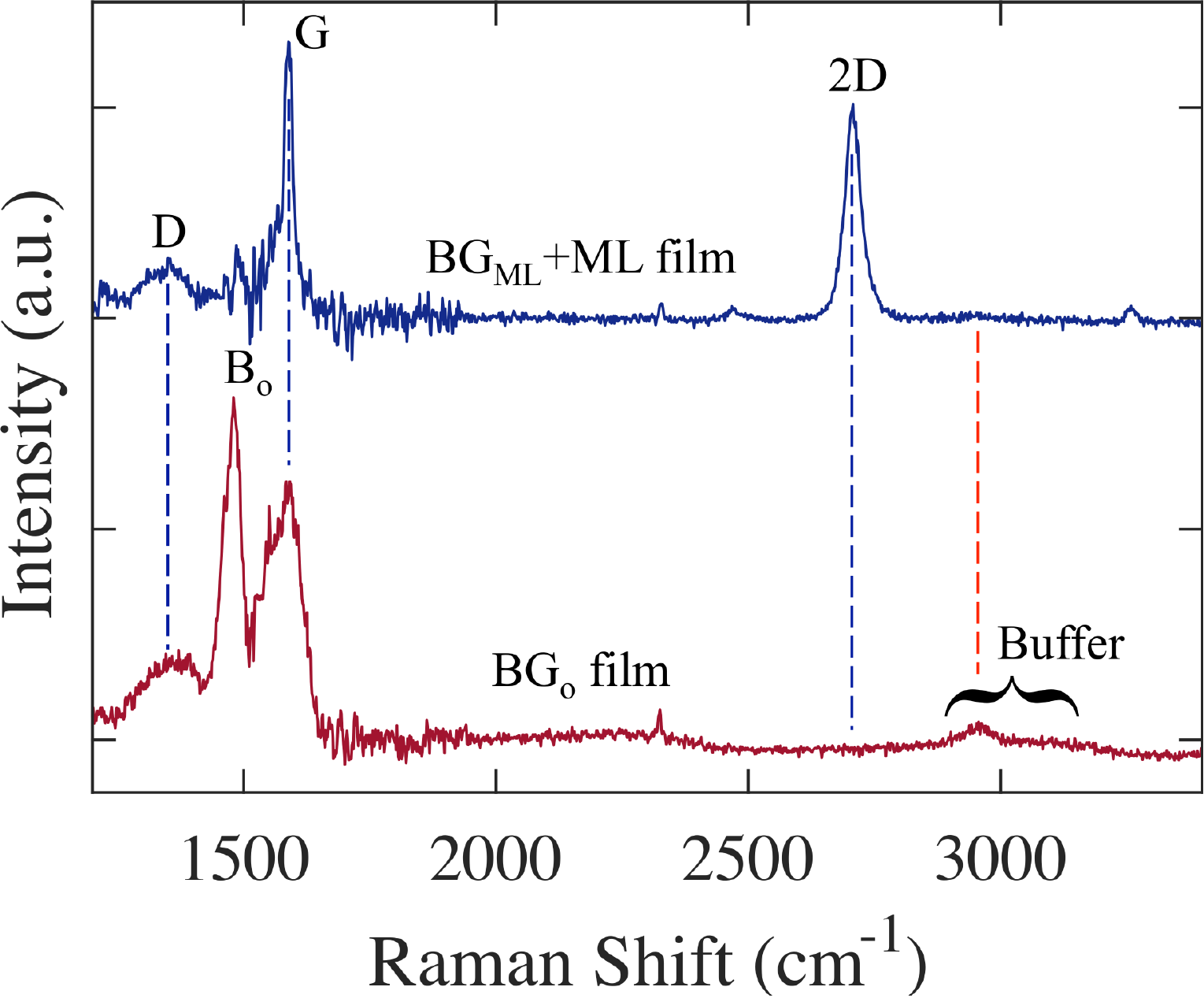}
\caption{A comparison of the Raman spectra from a buffer-only film and a ML graphene film grown on SiC(0001). A bulk SiC spectrum has been subtracted from each spectra. The ML film shows the D, G, and 2D peaks associated with graphene on SiC(0001).\cite{Kunc_APL_13} The \BGo~film lacks the 2D and sharp G peaks but has two characteristic features; the B\textsubscript{o} peak at 1480~cm\textsuperscript{-1} and a broad ``Buffer'' spectrum between 2900-3100~cm\textsuperscript{-1}. The ML coverage in the \BGo~film is less than 3\%.}
\label{F:Raman}
\end{figure}

Figure \ref{F:Raman} compares the Raman spectra of the \BGo~film to the spectra from a \BGML+ML  film. The \BGo~film spectrum lacks the ML  2D and sharp G peaks but shows two additional features; a B\textsubscript{o} peak at 1480~cm\textsuperscript{-1} and a broad G+D peak 
between 2900-3100~cm\textsuperscript{-1}.  The latter two features are known to be associated with a pure buffer film.\cite{Fromm_NJP_13,Palmer_APL_14} Using the ML 2D intensity as a reference and the background noise as an upper limit on the buffer 2D peak, we estimate that the ML coverage must be less than $<\!3\%$.  This is consistent with ARPES estimates of the ML coverage in a \BGo~film.\cite{Nevius_PRL_15}  The majority of the ML coverage is expected to be associated with ML that nucleates at intrinsic step edges.\cite{Emtsev_NM_09}  The Raman spectrum was measured at three positions (each 3~mm apart) on the sample and no significant changes in the Raman were found, indicating the large scale film uniformity necessary for XRR experiments. 

\subsection{XSW and XRR Methods}
 XRR  measurements were conducted at room temperature in UHV at the SIXS beamline at the Synchrotron SOLEIL  using a photon energy of $E\!=\!12.8$~keV. Prior to X-ray exposure, the samples were heated to 500\degC in UHV to remove absorbed contaminants.  The momentum transfer vector, ${\bf K}$, is defined as ${\bf K}\!=\!{\bf k}_f\!-\!{\bf k}_i$ where ${\bf k}_i$ and ${\bf k}_f$ are the momenta of the incident and reflected x-ray beams, respectively, For elastic scattering considered here,  $|k_f|\!=\!|k_i|\!=\!E/\hbar c$.  ${\bf K}\!=\!\left(h,k,l\right)$ represents a point in reciprocal space that is written in terms of the bulk hexagonal coordinates of SiC: ${\bf K}\!=\!h{\bf a}^*_{\text{SiC}}\!+\!k{\bf b}^*_{\text{SiC}}\!+\!l{\bf c}^*_{\text{SiC}}$ where $|{\bf a}^*_{\text{SiC}}|\!=\!|{\bf b}^*_{\text{SiC}}|\!=\!4\pi/\left(a_{\text{SiC}}\sqrt{3}\right)\!=\!2.3556\text{\AA}^{-1}$ and $|{\bf c}^*_{\text{SiC}}|\!=\!2\pi/c_\text{SiC}\!=\!0.6233\text{\AA}^{-1}$. The XRR polarization and geometric corrections were performed to allow comparison between the measured reflectivity intensities and calculated intensities from the model discussed in Sec.~\ref{S:Xray}.\cite{Vlieg_JAC_97} 

 In the XSW experiment, the sample is oriented so that the incoming x-ray beam of energy $E_B$ satisfies the condition that $\bf{K}\!=\bf{G}$, where $\bf{G}$ is a reciprocal lattice vector of SiC (i.e., a Bragg reflection). This geometry produces a standing wave whose crests are $\lambda\!=\!d\!=\!2\pi/|\bf{G}|$ apart and perpendicular to $\bf{G}$ [see Fig.~\ref{F:XSW_SCH}]. By changing the incident photon energy relative to $E_B$, the standing wave will move in the $\bf{G}$-direction exciting photoelectrons as a wave maxima passes through a plane of atoms [see Fig.~\ref{F:XSW_SCH}].\cite{Zegen_SSR_93,Bedzyk_2005}  The photoelectron intensity for a particular element $j$ in the material, $I^G_j(E)$, will therefore be a function of $E$ and thus related to the vertical position of that element. 

The XSW experiments were carried out at the GALAXIES beamline on the Synchrotron SOLEIL.\cite{Rueff_Galax_15}  We used the ${\bf G}\!=\!(004)$ Bragg reflection to produce the standing wave with $d\!=\!2.52~$\AA. The Bragg angle for the (004) reflection was $\theta_{B}\!\sim\!78^\circ$ corresponds to an incident photon energy of $E_B\!=\!2512$ eV.  All measurements were done at room temperature with an overall energy resolution better than 250~meV.  The take-of-angle, $\alpha$, for the photoelectron detector was $10^\circ$ to improve surface sensitivity in the XPS spectra. 
\begin{figure}
\includegraphics[angle=0,width=5.0cm,clip]{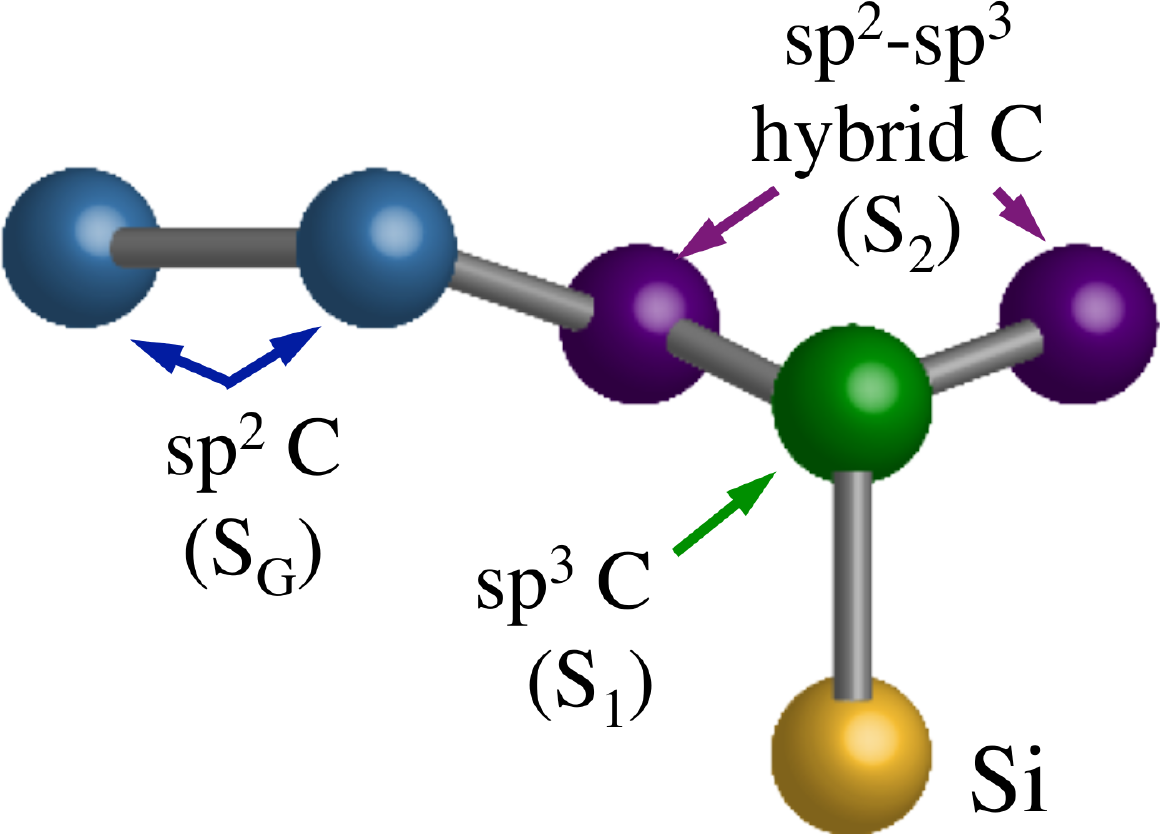}
\caption{ Schematic model of the transition from sp\textsuperscript{3} buffer carbon bonded to silicon in the SiC interface to sp\textsuperscript{2} graphitic carbon through a hybrid carbon back bond.  \Sone, \Stwo, and \SG~refer to the three C1s XPS components described in Sec.~\ref{S:results}}
\label{F:Buffer_bonding}
\end{figure}

For the studies presented in this work, we are interested in the photoelectron yield from the C1s core levels that have components from both the carbon in SiC and the carbon in the buffer layer. The details of analyzing the C1s spectra as a function of both photon energy $E\!-\!E_B$ and binding energy $BE$ require some care. We do this by first setting the number of components in the spectrum [see Sec.~\ref{S:XPS}] and then fitting each spectrum $I_E(BE)$ for a range of photon energies around $E_B$.  The fitting is done by minimizing the global $\chi^2_g$ [see supplemental material]. The global $\chi^2_g$ is the $\chi^2$ for a single C1s spectrum fit, at a single  photon energy, summed over fits for every incident photon energy in the data set.  While each component's line shape parameters, i.e., width and position, are allowed to vary, they are constrained to be independent of the incident photon energy.  Only the components' areas are allowed to vary as a function of $E$. A Shirley background has been subtracted from every core-level spectrum. We use a Doniach-Sunjic lineshape for the graphene peak and a Voigt lineshape for the buffer carbon, bulk carbon component, and Si 2s core-levels.
 
\section{Results }\label{S:results}
While XRR is able to study surfaces with buried interfaces, the phase problem makes structural measurements difficult. Emery et al.\cite{Emery_SW_13} have shown how combining XSW and XRR studies of the SiC-graphene interface can help overcome the phase problem. However, in the particular case of the buffer layer, XSW analysis has its own problems that were not recognized and that we now discuss before presenting our experimental results.
\begin{figure}
\includegraphics[angle=0,width=7.0cm,clip]{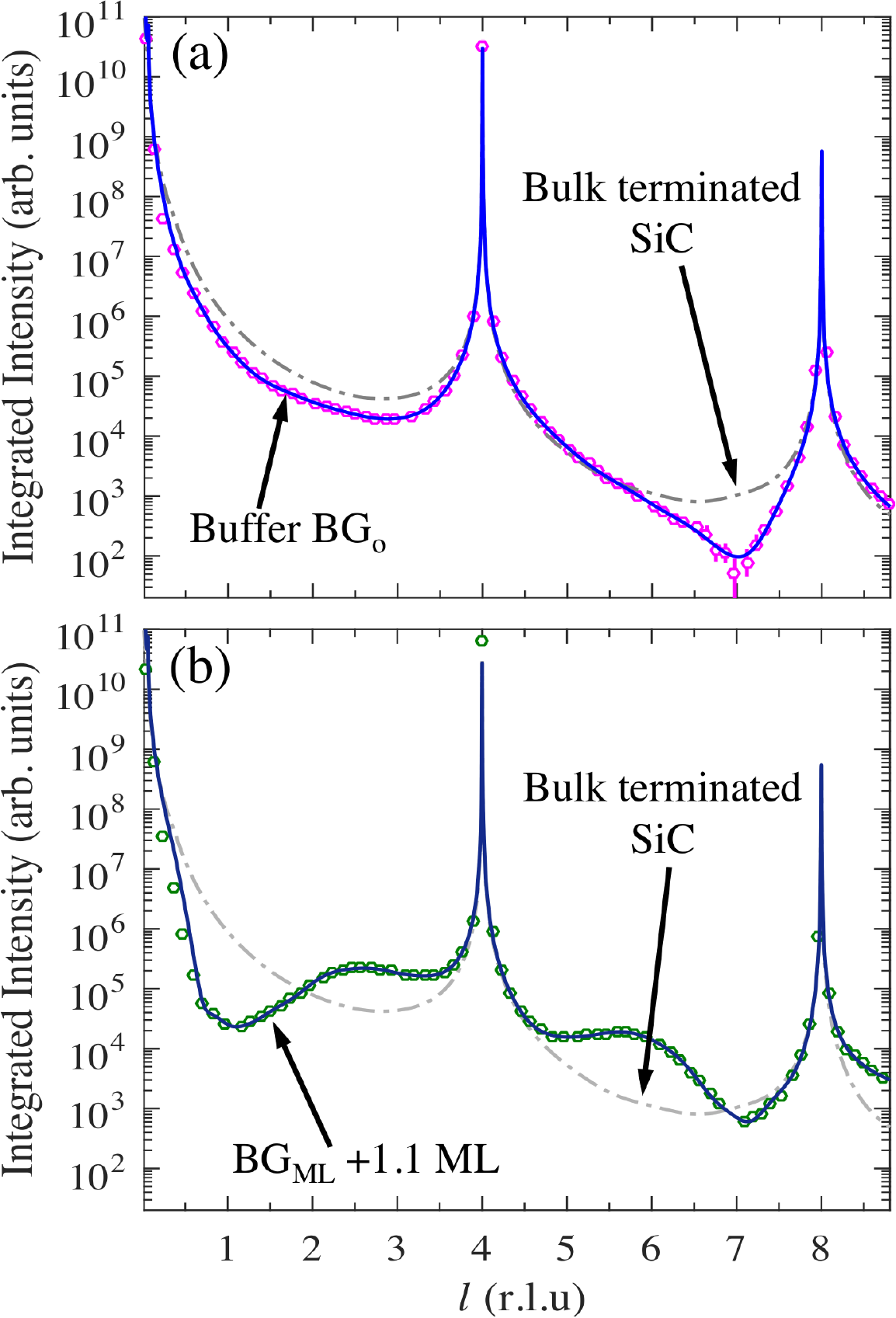}
\caption{The experimental X-ray reflectivity (open circles) of; (a) a \BGo~only film, and (b) a ML film above the \BGML~layer.  For the \BGo~film, the calculated reflectivity (solid lines) in (a) use a buffer comprised of three components (\Sone, \Stwo, and \SG) plus a small 1\%~ML coverage. Details of the calculation are given in Sec.~\ref{S:Xray}.  For the ML film in (b), the calculations uses a single \Sone~component buffer for the \BGML~layer.  For comparison, the calculated reflectivity from a bulk terminated surface (dashed lines) is shown in both plots. {\color{red} *** add a figure with XSW positions **}}
\label{F:XRef_Buffer_ML}
\end{figure}
\begin{figure*}
\includegraphics[angle=0,width=16cm,clip]{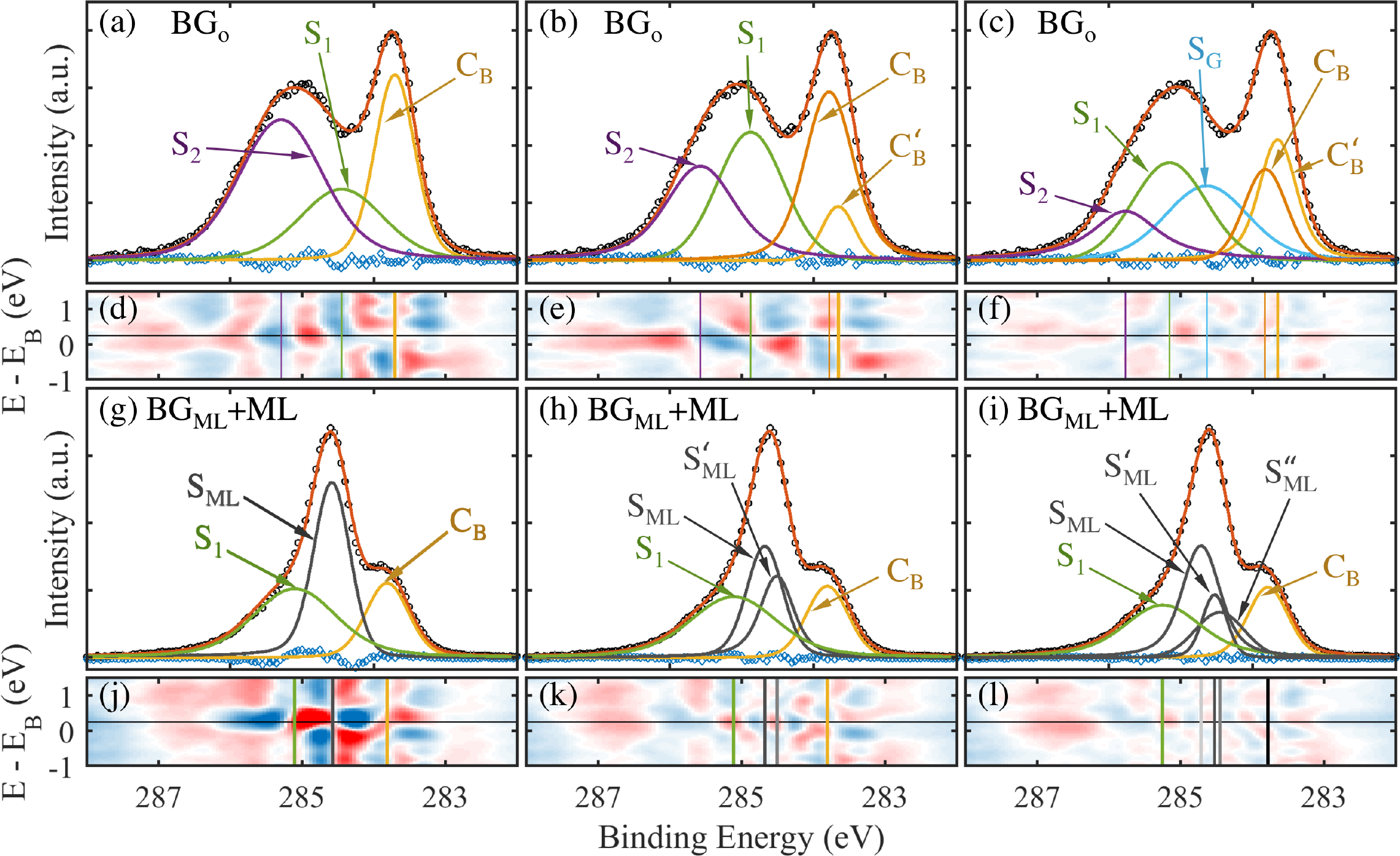}
\caption{The C1s spectral decomposition for a \BGo~film [(a), (b), and (c)] and a \BGML+ML [(g), (h), and (i)] films. The fits are at a photon energy of $E\!-\!E_B\!=\!0.25$~eV. The residual maps [(d), (e), (f), (j), (k), and (l)] for each fit are shown as a function of $E\!-\!E_B$ and BE. Vertical lines in the maps mark the best fit BE peak positions of each component. (a) A buffer-only C1s spectra using a 3-component fit with two buffer components, \Sone~and \Stwo~plus a bulk SiC component $\text{C}_\text{B}$. (b) A 4-component fit similar to (a) but with an additional component labeled $\text{C}'_\text{B}$. (c)  A 5-component fit similar to (b) but with the additional component S\textsubscript{G}. (g) A \BGML+ML film C1s spectra using a 3-component fit with a \Sone~buffer component, a ML component $\text{S}_\text{ML}$, and a bulk SiC component C\textsubscript{B}.  (h) A 4-component fit similar to (g) but with an additional component labeled $\text{S}'_\text{ML}$. (i) A 5-component fit similar to (h) but with a third component labeled $\text{S}''_\text{G}$.}
\label{F:XPScomp}
\end{figure*}
The C1s XPS spectra of graphene grown on SiC(0001) was thought to contain four components; \Sone, \Stwo, \SML, and \CB. \Sone~and \Stwo~are from the buffer carbon layer that contains C-Si bonds to the interface and distorted C bonds in the unbonded part of the buffer layer.  The \SML~component is from sp\textsuperscript{2} C-C bonds in any graphene that grows above the buffer. The \CB~component is from sp\textsuperscript{3} C-Si bonds in bulk SiC.\cite{Emtsev_PRB_08} However, several theoretical calculations along with experimental scanning transmission electron microscopy (STEM) and x-ray diffraction measurements show an incommensurate structure that suggests reduced interfacial bonding. All these works indicate that part of the buffer should also have sp\textsuperscript{2} bonded carbon.\cite{Kim_PRL_08,Conrad_NL_17,Nair_NL_17}  
This possibility is shown schematically in Fig.~\ref{F:Buffer_bonding} and implies at least one additional component (we label \SG) may be present in the buffer's C1s spectrum.  Any partial ML film would have an \SML~component obscuring any possible sp\textsuperscript{2} bonded carbon in the buffer layer's C1s spectrum. Furthermore, any substantial reconstruction of the the top SiC bilayer could also cause new components to arise in the C1s spectrum near the BE of the bulk \CB~component.  It will therefore be important in our subsequent XPS analysis to explore how additional C1s peaks affect the XSW results.

The large ML coverages in samples used in earlier studies cause an additional and more important problem. If the \BGo~and \BGML~layers were structurally identical, the ML coverage would only complicate the spectral analysis of the C1s spectrum.  We now know that this assumption is wrong and that there are substantial changes in the buffer's structure when a ML grows above it.\cite{Conrad_NL_17}  
This means that a substantial ML coverage will cause both the XSW and XRR data to give results that represent an unknown weighted average of two different structures making it impossible to determine a unique \BGo~structure.  For XRR, the ML coverage has a further complication. Because both the \BGo~and \BGML~layer distributions are broad in $z$ while ML or bilayer graphene layers are narrow,\cite{Emery_SW_13,Sforzini_PRL_15} the buffer's contribution to the scattered amplitude will be smaller than the ML's.\cite{Robinson_SS_92} This effect is demonstrated by comparing the calculated XRR reflectivity of a bulk terminated SiC surface to the measured buffer or ML films
[see Fig.~\ref{F:XRef_Buffer_ML}].  Between Bragg peaks ($l$= 0,4, and 8), the buffer layer causes small changes (about a factor of 4-10) with respect to the bulk terminated surface.  A ML film on the other hand can produce changes that are a factor of 10 larger and thus obscure reflectivity contributions from the buffer film [see Fig.~\ref{F:XRef_Buffer_ML}(b)].

The problems associated with ML films we identified above are the reasons we use samples with high coverage control. Our XSW studies will consider the possibility of new components in the C1s fits.  Recognizing the possible systematic errors associated with multiple peak fits in the C1s spectra, we will use the XSW derived $z$-distributions as starting parameters for XRR reflectivity fits.  These distributions will be allowed to vary (with restrictions) to refine the  \BGo~and \BGML~layers structure.

\subsection{Fitting the XPS Spectra}{\label{S:XPS}
Figure \ref{F:XPScomp}(a) shows a  3-component fit (\Sone, \Stwo, and C\textsubscript{B} used in previous studies) to the \BGo~layer when no ML graphene is expected on the sample. It must be emphasized that the relative areas of the individual components in Fig.~\ref{F:XPScomp} are not a measure of their relative concentrations as will be discussed in Sec.~\ref{S:XSW}.  In fact the photon energy used for the C1s spectra shown in Fig.~\ref{F:XPScomp} was deliberately chosen to minimize the \Sone~component that can obscure any components that occur in an energy range where a graphene ML component would appear.
Our 3-component fit is very similar to those found in the work of Emtsev et al.\cite{Emtsev_PRB_08}. In that earlier work, the C1s spectrum was monitored as a function of growth, so we would expect very little ML coverage in their early growth spectra, explaining the similarity of their fit to ours.  
 
 While 3-components can fit the C1s spectrum, there are large residuals [see Fig.~\ref{F:XPScomp}(d)]  suggesting that other peaks could be present. As discussed above and shown in Fig.~\ref{F:Buffer_bonding}, we expect the \BGo~layer to have a structure that could have an additional C1s component. 
However, if we add a 4\textsuperscript{th} component near the BE of ML graphene (285.5~eV), the fitting procedure moves the peak to a position close to the bulk \CB~position.  Because of its proximity to \CB, we label the new peak \CBp~[see Fig.~\ref{F:XPScomp}(b)]. It is certainly reasonable that the top SiC bilayer would relax as some bilayer Si bonds to the buffer and distorts the bilayer Si-C bonds. This would strain the top SiC bilayer from the bulk configuration and lead to a range of binding geometries that would broaden the \CB~peak and possible give rise to a \CBp~component.
The overall effect on the C1s fit using the \CBp~component is very dramatic.  The intensity ratio of \Sone~to \Stwo~reverses and both peaks become narrower as \Sone~moves to higher BE. Note that despite these changes, the residual map is only slightly improved [see Figs.~\ref{F:XPScomp}(e)].

The residual map for the \BGo~film can be substantially improved by adding a 5\textsuperscript{th} component [see Figs.~\ref{F:XPScomp}(c) and (f)].  The best fit position of the new peak is labeled \SG~because its BE is near that of graphite, between the BE of the buffer peaks and the bulk carbon peaks. The \SG~peak causes \Sone~to move to even higher BE and further increase the intensity ratio of \Sone~to \Stwo.  In fact, the spectral ratio of \Sone~to \Stwo~in the 5-component fits resembles that found by Emery et al.,[\onlinecite{Emery_SW_13}] using a ML film.  The similarity between our 5-component fit and Emery et al.,[\onlinecite{Emery_SW_13}] might be expected given that the \SG~peak influences the fit shape in much the same way as a true ML graphene peak would.  The peak positions and widths of the different component fits for the \BGo~C1s spectra are summarized in Table \ref{T:XPS}.  As we'll show in Sec.~\ref{S:XSW}, the three different C1s decompositions lead to very different \BGo~layer structures.  
\begin{table}[htpb]
\caption{\label{T:XPS}The fitted C1s BE positions and FWHM (FW$_\text{j}$) for both buffer-only and buffer+ML films. BE is given as shifts, $\Delta E_j$, relative to the bulk carbon BE at 283.7~eV.  All values are in eV and the statistical uncertainty is 0.1eV. }
\centering
\begin{tabular}{l  l| c c|  c c|  c c|  c c| c }
\hline \hline
&&$\Delta E_{S2}$ &FW$_\text{S2}$ &  $\Delta E_{S1}$ & FW$_\text{S1}$ &$\Delta E_{SG}$& FW$_\text{SG}$\\ 
\hline
\multirow{3}{*}{Buffer} 
&3-Comp & 1.58    &	1.38			& 0.74    			&1.38 		&-  				&-	 \\
&4-Comp & 1.92    &	 1.15			& 1.23    			& 1.07   		&-				&- 	 \\
&5-Comp & 2.12    &	 1.12			& 1.51   	 		&1.21		&0.99 			&1.37 \\
\hline\hline
&		&	    &	     			&  $\Delta E_{S1}$ 	& FW$_\text{S1}$ &$\Delta E_{ML}$& FW$_\text{ML}$\\ 
\hline
ML 
&3-Comp	&-	     &-   			&1.35			& 1.37  		    &0.77			& 0.57	\\
 \hline   \hline
\end{tabular}
\end{table}

There are also multiple ways to fit the ML film's C1s spectra. Figures \ref{F:XRef_Buffer_ML}(g-i) show fits for a \BGML+ML film using different numbers of C1s components. The 3-component fit [Figs.~\ref{F:XRef_Buffer_ML}(g)] gives a reasonable fit to the ML C1s spectrum. Note that we label the higher binding energy peak as \Sone~not \Stwo.  This is because its BE lies closer to \Sone~in the \BGo~4- and 5-component fits.  As we'll show in Sec.~\ref{S:XSW}, the assignment of the highest BE peak to \Sone~will be supported by the XSW results.  Adding other components significantly improve the residuals [see Figs.~\ref{F:XRef_Buffer_ML}(j), (k), and (l)] but has remarkably little affect on the parameters for the \Sone~component. We label these additional peaks $\text{S}'_\text{ML}$ and $\text{S}''_\text{ML}$ because no matter the starting position, their best fit BE cluster around the \SML~monolayer peak's BE. While adding $\text{S}'_\text{ML}$ and $\text{S}''_\text{ML}$ improves the fits, they are too weak and too close together to get reliable yield curves for subsequent XSW analysis [see supplemental material].  This is in part due to attenuation effects of the \BGML~components by the ML film above.  We will therefore only report on the \Sone~and \SML~parameters for the 3-component fits parameters [see Table~\ref{T:XPS}] and wait until the XRR data is presented to give a better estimate of the \BGML~structure.  

\subsection{X-ray Standing Waves}{\label{S:XSW}
The normalized photoelectron yield $Y^G_j(E)$ for element $j$ in an XSW experiment is given by:\cite{Zegen_SSR_93,Bedzyk_2005}
\begin{equation}
  \begin{aligned}
Y^G_j(E)=1+& R^G(E)+ 2\sqrt{R^G(E)}\times\\ 
& \int_{vol} \rho_j'(r)\cos(\nu(E)-2\pi P^G_j)dr.
 \end{aligned}
\label{E:SW}
\end{equation}
$R^G(E)$ is the energy dependent x-ray reflectivity at the Bragg point ($G\!=\!\text{(004)}$ in this work) and $\nu (E)$ is the phase of the standing wave [see supplemental material].\cite{Zegen_SSR_93} Both are calculated from the bulk crystal structure.\cite{Authier_2001}  $\rho_j'(r)$ is the density distribution of the $j^\text{th}$ element and $P_j^G=\bf{G}\cdot\bf{r_j}/(2\pi)$ [where $r_j$ is the position of the j\textsuperscript{th} atom, modulo $d$]. Rather than deal with arbitrary atom distributions, we will only consider atoms vertically distributed in discrete planes at positions $z_j$ in a Gaussian distribution.   This is done by writing $\rho_j'(r)$ as a convolution of a delta function with a normalized gaussian distribution whose width is $\sigma_{Z,j}$.  This reduces Eq.~\ref{E:SW} to:
\begin{subequations}
\begin{equation}
\begin{aligned}
Y^G_j(E)=&1+ R^G(E)+\\
&  2\sqrt{R^G(E)}f_j\cos(\nu(E)-2\pi P^G_j).
 \end{aligned}
\label{E:SW_final}
\end{equation}
\begin{equation}
f_j= e^{-\sigma_{Z,j}^2G^2/2}.
\label{E:SW_fj}
\end{equation}
\end{subequations}
$Y^G_j(E)$ is related to the experimental intensity by a normalizing constant; $ I^G_j(E)\!=\!N_jY^G_j(E)$.  Neglecting photoelectron attenuation (a reasonable assumption for the buffer only film) the coverage of a given C1s component is then related to the $N_j$'s; $\theta_j\!\sim\!N_j/\Sigma N_j$

\begin{figure}
\includegraphics[angle=0,width=8.4cm,clip]{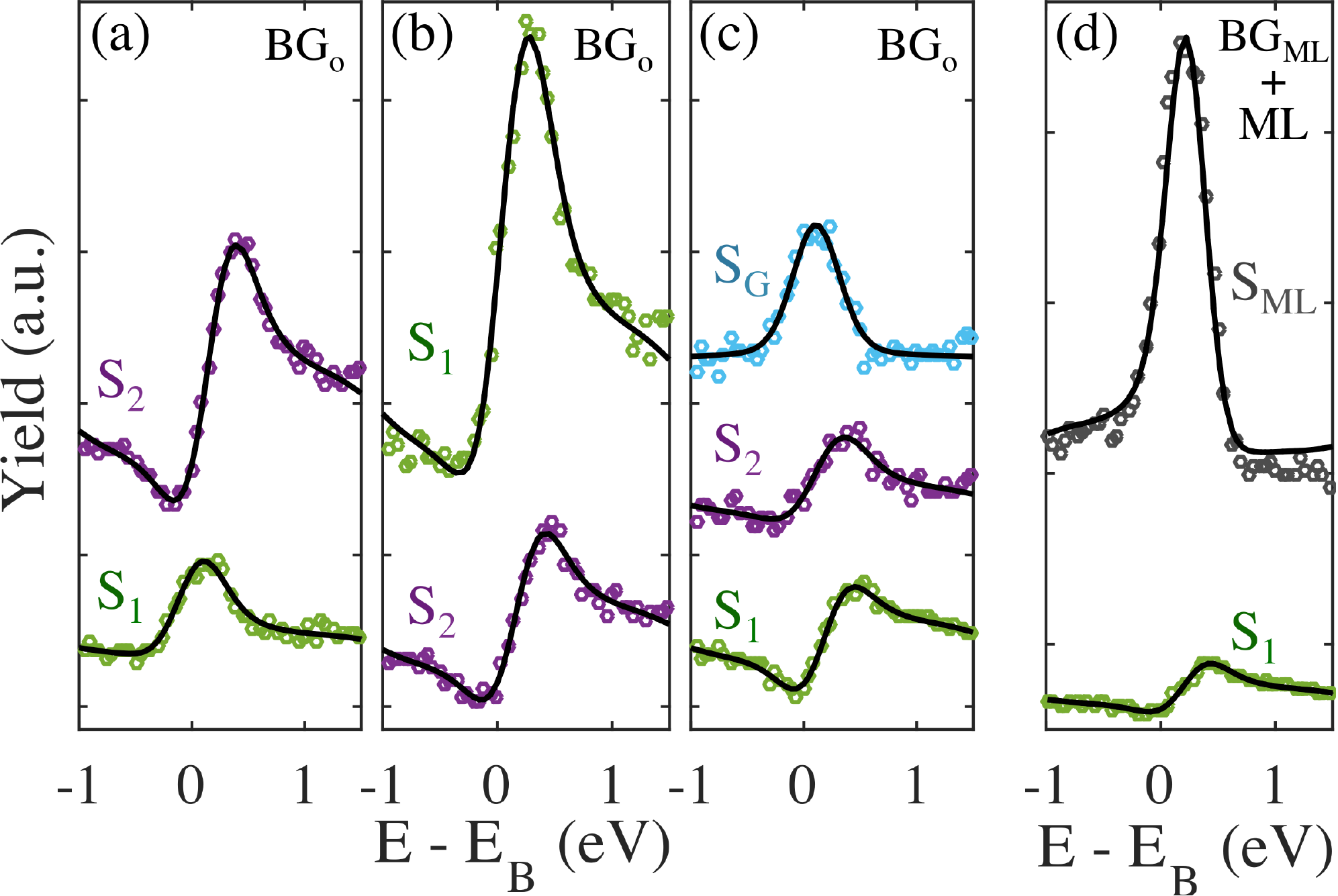}
\caption{(a), (b), and (c) show XSW yield data ($\circ$) for the \BGo~film using 3-, 4-, and 5-components fits for the C1s spectra, respectively. Solid lines are fits using Eq.~\ref{E:SW_final}. (d) The yield data and fit for a \BGML+ML film derived from a 3-component C1s fit.}
\label{F:Yield}
\end{figure}

\begin{figure}
\includegraphics[angle=0,width=6.7cm,clip]{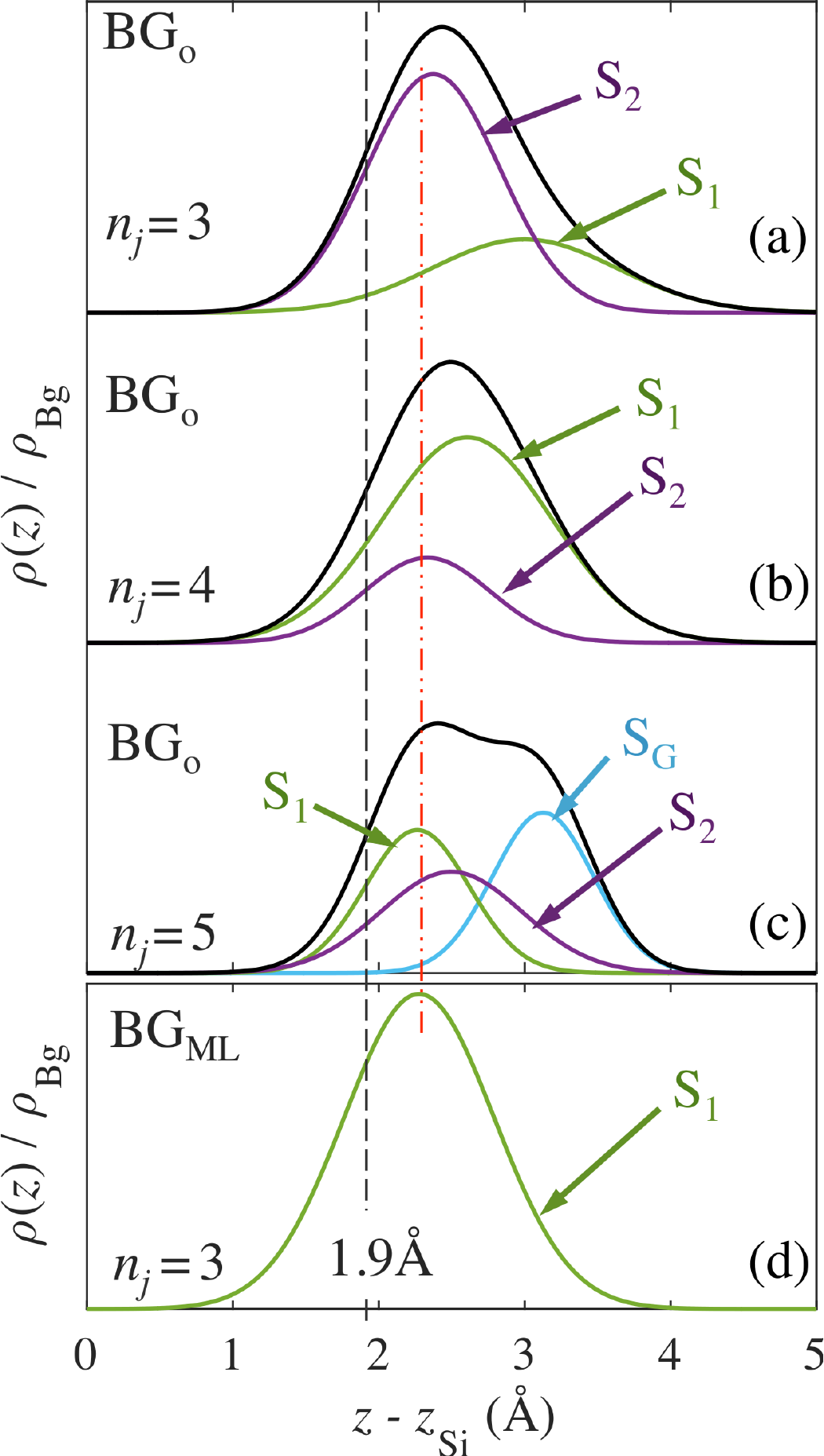}
\caption{Comparison of the XSW derived vertical density distributions, $\rho(z)$, of the \BGo~and \BGML+ML films relative to the last Si layer in the SiC(0001) surface (located at $z_\text{Si}$). $\rho(z)$ is normalized to the \BGo~density. (a), (b), and (c) are for the \BGo~film using 3-, 4-, and 5-component fits, respectively. (d) is the profile of the \BGML~layer using 3-components. The red vertical dashed-dot line marks the \Sone~position in the \BGo~film in (c). The black vertical dashed line marks the distance between buffer-carbon and Si measured by STEM.\cite{Nair_NL_17}}
\label{F:XSWz}
\end{figure}
Figures \ref{F:Yield}(a), (b), and (c) show the experimental yields for the \BGo~film derived from 3-,4-, and 5-component C1s fits. The fit for the ML film, using 3 C1s components, is shown in Fig.~\ref{F:Yield}(d).  All fit parameters are summarized in Tables \ref{T:XSW_Buffer} and \ref{T:XSW_Bulk}. While all the \BGo~fits are very good, regardless of the number of components in the spectral decomposition, there are major differences in their $z$-distributions. To make comparisons between the different fits easier to visualize, we have plotted their different $z$-distributions in Fig.~\ref{F:XSWz}.  For reference, we have marked the 1.9\AA~C-Si bond distance between buffer carbon and interfacial silcon measured by STEM in Fig.~\ref{F:XSWz}.\cite{Nair_NL_17} Note that the STEM value is close to the 1.89\AA~bulk SiC bond length. 

There are a number of observations from the XSW results that indicate that the 5-component fit is closer to the actual \BGo distribution. First, the 3- and 4-component fits give a \BGo~distribution that is essentially a broad Gaussian with the \Stwo~component closer to the SiC interface [Figs.~\ref{F:XSWz}(a) and (b)]. The essentially Gaussian shape is \emph{inconsistant} with experimental STEM profiles of the buffer that indicate a sharper $z$-distribution.\cite{Nair_NL_17}  In fact the 3-component fit finds that the \Stwo~atoms closer to the SiC are more than twice as numerous than the \Sone~atoms farther away [see Table \ref{T:XSW_Buffer}]. This contradicts both STEM results and theory predictions that the bonded carbon (presumably \Stwo~in the 3- and 4-component fit) is a much smaller fraction of the buffer carbon.  Finally, both the 3- and 4-component fits find the unphysical result that 23\% of the buffer carbon lies closer to Si than the smallest known C-Si bond distance of 1.89\AA.

\begin{table}[htpb]
\caption{\label{T:XSW_Buffer}XSW results for the \BGo~and \BGML+ML films.  Results are shown for 3-, 4- and 5- componets fits to the \BGo~film. The parameters for the ML film only include results from the 3-component fit.  The $Z$ positions are shown relative to the position of the last Si layer.}
\centering
\begin{tabular}{ c  c   | c | c| c| c| c}
\hline\hline
		 	&  & \multirow{2}{*}{$P$} & $(Z-Z_\text{Si})\footnotemark[1] $	& \multirow{2}{*}{$f$}	 & $\sigma$ & 																		\multirow{2}{*}{$\theta$} \\
			&  &					&	(\AA)						&				&(\AA)    &     \\ 			
\hline
\multirow{2}{*}{}{\BGo} 	&\Stwo &-0.02(2)  	& 2.37(4) 			    			& 0.52(4) & 0.46(2)&0.70(1) \\ 
                        (3-peak)	&\Sone & 0.23(5) 	& 3.01(1) 			    			& 0.3(4)   & 0.6(1)  &0.30(1)\\ 
\hline
\multirow{2}{*}{} \BGo	&\Stwo &-0.04(1) 	& 2.33(4) 			     			& 0.56(5) & 0.43(3) &0.24(1)\\ 
		       (4-peak)   	&\Sone &0.07(3)   	& 2.61(7) 			     			& 0.36(4) & 0.58(3)  &0.76(2)\\
\hline 
\multirow{2}{*}\BGo		&\Stwo &-0.03(5) 	&{\color{red}\textbf{2.5(1)}}		& 0.48(9) & 0.48(7) &0.32(1)\\ 
\multirow{2}{*}{(5-peak)}	&\Sone &-0.07(1) 	&{\color{red}\textbf{2.26(3)}}		& 0.67(5)& 0.36(4) &0.31(1)\\
     				&S\textsubscript{G} 	&0.28(2)  	& 3.13(6)   				& 0.7(2)  & 0.3(1)   &0.36(2)\\

    \hline \hline
\multirow{2}{*}{\BGML}  	&\Sone 		&-0.06(2)  & 2.30(5)  					& 0.41(2) & 0.52(3) &-\\ 
  					&ML			 &0.27(2)   & 5.63(5)  					& 1.0(1)  	& 0.0(1) &-\\     
    \hline\hline
\end{tabular}
\footnotetext[1]{The position of the Si layer is determined by the XSW yield fits to the Si 2s core level; $Z_\text{Si}\!=\!0.09$\AA~[see Table \ref{T:XSW_Bulk}].}
\end{table}
\begin{table}
\caption{\label{T:XSW_Bulk}XSW results for bulk carbon C1s and bulk silicon Si 2s.}
\centering
\begin{tabular}{l l | c | c | c  | c }
\hline\hline
		 		&  					& $P_j$ 	& $Z_j$ (\AA) 	& $f_j$ 		& $\sigma_j$ (\AA)\\
\hline
\multirow{2}{*}{\BGo} & \CB 	&-0.24(1) 	& 1.92(2) 		& 0.79(5) 		& 0.27(4) \\ 
				&Si\textsubscript{B} 	&0.04(1)  & 0.09(2)  		& 1.00(3) 		& 0.05(4)  \\ 
\hline\hline
\multirow{2}{*}{\BGML}	&\CB 	&-0.25(1)  & 1.89(2)  	& 0.72(5)  		& 0.33(2) \\ 
				&Si\textsubscript{B} 	&0.04(1)   & 0.09(2)  		& 0.99(3) 		& 0.07(4) \\ 
\hline\hline
\end{tabular}
\end{table}

There is one more result that points to the 5-component C1s fit being the correct deconvolution of the \BGo~XPS spectrum.  As already noted, XSW analysis predicts that the \Stwo~component from 3- and 4-component fits to the C1s XPS spectra fits is associated with buffer carbon closest to the last bulk Si layer. This was the same result found in XSW experiments by Emery et al.\cite{Emery_SW_13} using a ML sample.  However, our 5-component XSW analysis shows just the opposite result. Using the 5-component fit, we find that \Sone~is closest to the Si layer indicating that \Sone, not \Stwo, is from buffer carbon bonded to silicon. The proposal that \Sone~is closest to the Si layer was first made by Emtsev et al.\cite{Emtsev_PRB_08} in XPS studies on a series of samples ranging from no graphene to several monolayers.  They noted that the buffer's $\sigma$-bands were downshift by 1.0~eV compared to graphite.   The shift meant that the \Sone~and \Stwo~buffer carbon peaks were also shifted to high BE relative to neutral graphite. Since the sp\textsuperscript{2} bonded carbon must be at a higher BE than C-Si bonds, they concluded that the \Sone~feature must be associated with buffer-Si bonds. The Emtsev interpretation that \Sone~is closer to the SiC is further supported by comparing our ML and buffer layer results.  We only need  3-components to fit the \BGML+ML film's C1s spectra: the ML \SML~peak,  the bulk \CB~peak, and a component from the \BGML~layer.  The \BGML component was labeled as \Sone~in Fig.~\ref{F:XPScomp}.  Since the \BGML~\Sone~component is the only component close to the interface in the ML film, it must have a considerable fraction of carbon bonded to Si.  In other words, the BE of \Sone~in the ML film should be similar to the BE of the component in the \BGo~layer that is bonded to Si in the interface.  At the same time, the components that binds to Si in both the \BGo~and \BGML~must be approximately the same $z$-distance from the top Si layer.  A comparison of the \BGo~and \BGML+ML film data in Table \ref{T:XPS} and the red dashed line in Fig.~\ref{F:XSWz} shows that only the \Sone~peak in the 5-component C1s spectrum meets both these combined requirements. Based on these observations, we propose that the 5-component fit to the \BGo~layer's XPS data represents the actual C1s spectra and that the previous Emtsev et al.\cite{Emtsev_PRB_08} interpretation correctly assigns \Sone~as being the \BGo~carbon component bonded to Si. The incorrect assignment of the bonding species in the prior XSW analysis was most likely due to the XPS signal being a mixture of of two different spectra from ML and buffer layer partial coverages.

The XSW derived distance between the \Sone~carbon in the \BGo~layer and the top Si layer is 2.26\AA, comparable to the 2.1(1)\AA~distance measured in previous XSW work\cite{Emery_SW_13} but is larger that the 1.9\AA~bond length measured by STEM.\cite{Nair_NL_17} The differences between the XSW results and STEM values is likely due to assigning the Si top layer $z$-position as the zero of the $z$-scale based solely on the value derived from the Si~2s XSW yield.  Since the Si~2s yield contains contributions from several bilayers, which we will show have different $z$-distributions, the XSW derived Si position will have systematic errors. 

Finally, even though the 5-component gives the best overall density profile, their are problems.  The density profile is not as sharp as sharp as STEM results would have predicted. Furthermore, we expect from the model in Fig.~\ref{S:results} that there should be significantly more \Stwo~carbon compared to \Sone.  The XSW results in Table \ref{T:XSW_Buffer} give nearly the same concentrations for both \Sone~and \Stwo.  We believe these problems with the XSW results are associated with the difficulty of accurately deconvoluting the closely spaced C1s components in the XPS spectra.  To improve on the XSW structural result we must combine these findings with our XRR results of the \BGo~and \BGML+ML films.

\subsection{Surface X-ray Diffraction}\label{S:Xray}
To fit the x-ray data, we use four SiC bilayers above the bulk with a buffer graphene layer above [see Fig.~\ref{F:XSW_SCH}. We also allow up to two partial graphene layers above the buffer. The scattered x-ray intensity $I(\Theta,\ell)$ is then:
\begin{equation}
\begin{aligned}
I(\Theta,\ell)= & A(\Theta,\ell) e^{-4\gamma_{\text{SiC}}\sin^2{\pi \ell/2}}\left |\frac{F_\text{bulk}(\ell)}{1-e^{-2\pi{i\ell}}} \right.\\
&\left. + F_{I}(\ell) + \frac{\rho_{\text{G}}}{\rho_{\text{SiC}}}(F_\text{BG}(\ell)+F_\text{G}(\ell))\right |^2,
\label{E:intenisty}
\end{aligned}
\end{equation}
where $F_\text{bulk}$ is the bulk 4H-SiC structure factor,\cite{Bauer_ACryst_01} modified by the crystal truncation term, $(1-e^{-2\pi{i\ell}})^{-1}$ [see Ref.~\onlinecite{Robinson_CTR_86}], $F_{I}$ is the structure factor of the 4-bilayer SiC interface region , $F_\text{BG}$ is the buffer graphene structure factor, and $F_\text{G}$ is the structure factor of any ML graphene layers above the buffer.  $F_\text{BG}(\ell)$ and $F_\text{G}(\ell)$ in Eq.~(\ref{E:intenisty}) are weighted by the ratio of the areal atomic densities of a 4H-SiC$(0001)$ and a graphene (0001) plane; $\rho_{\text{G}}/\rho_{\text{Si}}$=3.132. The factor properly normalizes the scattered amplitude from the graphene layer per 4H-SiC$(0001)$ $(1\!\times\!1)$ unit cell.
$A(\Theta,\ell)$ is a term that contains all corrections due to the experimental geometry.\cite{Robinson_review,Vlieg_JAC_97,Feng_thesis} The exponential term accounts for the substrate roughness caused by half-cell step fluctuations in the SiC surface (the predominant step height on 4H samples;\cite{Hass_APL_06} $c_\text{SiC}/2$). $\gamma_{\text{SiC}}$ is the variance in the number of half-cell layers in the surface due to steps.\cite{Elliott_PYSICAB_96} Roughly, $\gamma_{\text{SiC}}$ is proportional to the SiC step density.

$F_{I}(\ell)$ in Eq.~(\ref{E:intenisty}) is the structure factor of the top four SiC C-Si bilayers plus an additional layer of Si to allow for the possibility of Si adatoms or a relaxed Si layer bonded to the buffer graphene [see Fig.~\ref{F:XSW_SCH}]. The interface structure factor is then:
\begin{equation}
F_{I}(\ell)=\sum_{j=1}^9{f_j(\ell )\rho_j   e^{-\sigma_j^2   (2\pi\ell/c_\text{SiC})^2   /2}     e^{i2\pi\ell z_j/c_\text{SiC}}},
\label{E:Surf_form}
\end{equation}
where $\rho_j$ is the relative atom density for the $j^{\text{th}}$ interface layer at a vertical position $z_j$ ($\rho_j=1$ for a bulk layer corresponding to $8.22\!\times\!10^{-16}\text{atoms/cm}^{2}$). It was found that the additional Si layer was not needed to fit the experimental reflectivity and will not be discussed in Sec.~\ref{S:Disc}. The zero height is chosen as the top layer of Si atoms in the top SiC bilayer. $f_j(\ell )$ is the atomic form factor of C or Si depending on the layer.  A normalized Gaussian of width $\sigma_j$ has been convoluted with each layer to included possible layer disorder (similar to Eq.~\ref{E:SW_fj} used to describe the XSW vertical distribution).

To be consistent with the XSW results, we allow the buffer to be composed of $n$ distinct carbon layers. The multilayer graphene structure factor can then be written in a general form similar to Eq.~\ref{E:Surf_form}:
\begin{equation}
F_{BG}(\ell) = f_{C}(\ell )P_{BG}\sum_{s=1}^{n}\theta_se^{-\sigma_s^2   (2\pi\ell/c_\text{SiC})^2   /2}  e^{2\pi{il}z_s/c_\text{SiC}},
\label{E:Fpb}
\end{equation}
where $f_{C}$ is the atomic form factor for carbon, $P_{BG}$ is the areal coverage of buffer graphene, and $\theta_s$ is the fractional coverage of each component in the buffer such that $\Sigma \theta_s=1$.  The structure factor of graphene above the buffer layer is:
\begin{equation}
F_{G}(\ell) = f_{C}(\ell )\sum_{m=1}^{M}P_{m}e^{-\sigma_m^2   (2\pi\ell/c_\text{SiC})^2   /2}  e^{2\pi{il}z_m/c_\text{SiC}},
\label{E:FML}
\end{equation}
where $P_{m}$ is the coverage of the m\textsuperscript{th} graphene layer.

To fit the experimental reflectivity to Eq.~\ref{E:intenisty}, we use a lasso fitting routine.\cite{Lasso}  The lasso technique allows us to use the XSW derived buffer and known SiC bulk parameters as starting points (default parameters).  Changes in these parameters are penalized in the ordinary least squares regression (OLS).  The penalties are initially set to be very large to identify which parameters in Eq.~\ref{E:intenisty} give the largest reduction in the OLS $\chi^2$. Gradually, all penalties are reduced and new default parameters are updated until the model has converged to the minimized $\chi^2$ [see supplemental material].\cite{MConrad_thesis} This enables a seamless connection of the XSW derived parameters with the XRR parameters, placing the results of both techniques on an equal footing. 
  \begin{table*}[htpb]
\caption{\label{T:SXRD}Comparison of the \BGo~and \BGML~structural parameters from both XRR and XSW analysis.  $\Delta_\text{Si-C}$ is the C-Si vertical separation in the top SiC bilayer [see Fig.~\ref{F:XSW_SCH}].  $\Delta_\text{Si-C}$ for the top bilayer is measured directly in XRR, but represents a more bulk-like value in XSW results.}
\centering
\begin{tabular}{ cc | ccc | ccc | ccc | c |
cc |}
\hline \hline

	&&&\multirow{3}{*}{$Z$-$Z_\text{S1}$(\AA)}
	&&& \multirow{3}{*} {$\sigma$(\AA) } 
 	&&& \multirow{3}{*} {$\theta$(ML\textsubscript{G}) } 
 	&&\multicolumn{3}{c|}  {Interface layer} 
 			 \\

	&	&	&	&	&	&	&	&	& 	&&$\Delta_\text{Si-C}$ 
	&\multicolumn{2}{c|}  {$\theta$ (ML\textsubscript{SiC})}
\\
 			&& \Sone & \Stwo & S\textsubscript{G}
 			& \Sone & \Stwo & S\textsubscript{G}
 			& \Sone & \Stwo & S\textsubscript{G}
 			
			& (\AA) 
			&  Si & C
 \\ \hline

\multirow{2}{*}{\BGo}

& XRR	&1.9(1) &2.7(1)  &3.8(1)    &0.15(5)  &0.27(2)  &0.3(1)    &0.26(6) & 0.47(6)  &0.26(4)  &0.47(2) 
& 0.75(10) & 0.9(2) \\
& XSW	&2.26(3) &2.5(1)  &3.13(6)    &0.36(4)  &0.48(7)  & 0.3(1)    &0.33(1) &0.32(1)&0.36(2)  &0.69(2)
& - & -\\

\hline\hline
\multirow{2}{*}{\BGML}	
& XRR	&2.22(7)   	 &-	 & -  	&0.2(1)	&-    &-    &0.9(1) &-   &-   & 0.46(5)	
& 0.79(5) &0.8(1)   \\
& XSW	& 2.30(5) 	& -  	& -    &0.52(3)  & -   & -    &- 	   &-   &- 	
& 0.33(2) 	& 
& -\\
  \hline   \hline
\end{tabular}
\end{table*}

\begin{figure}
\includegraphics[angle=0,width=8.2cm,clip]{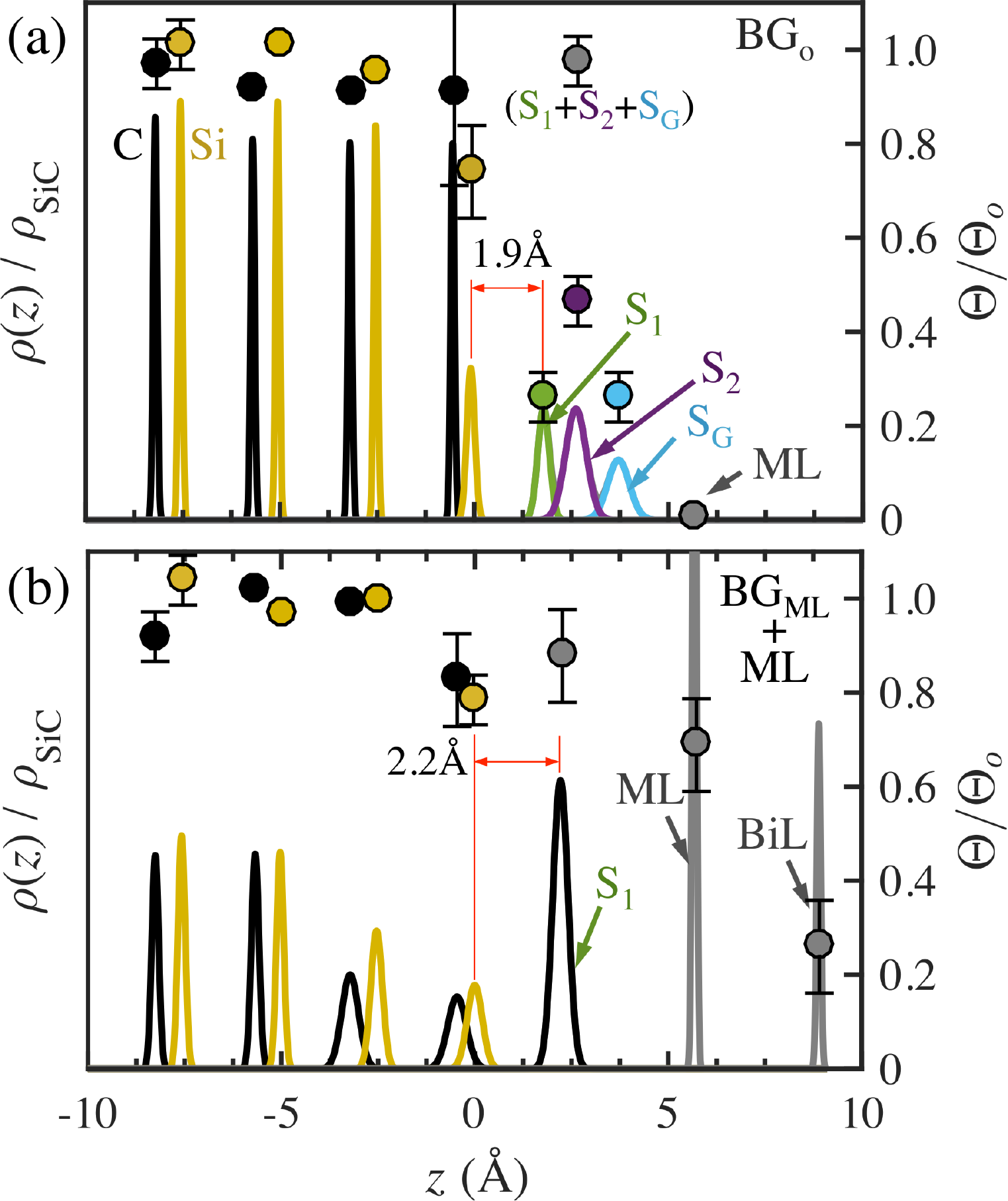}
\caption{The XRR derived vertical density profiles, $\rho (z)$, for both the \BGo~(a) and the \BGML+ML (b) films.  Densities are normalized to the bulk SiC density.  Solid lines are the density profiles (left axis) for SiC Si  (yellow) and C atoms (black), for buffer \Sone~(green), \Stwo~(purple), \SG~(cyan) carbon species, and for ML or bilayer graphene (grey).  Circles represent the integrated layer coverages, $\theta$, of each layer or component (right scale). $\theta$ for the bulk components are relative to the SiC in-plane density while the buffer and ML $\theta$s are relative to a ML graphene density.}
\label{F:Z-Profiles}
\end{figure}

\section{Discussion}\label{S:Disc}
Figure~\ref{F:Z-Profiles} shows a comparison of the density profiles from the buffer \BGo~and \BGML+ML films derived from the XRR fits in Fig.~\ref{F:XRef_Buffer_ML}.  The fit parameters are given in Table \ref{T:SXRD}. The calculated reflectivity for a buffer \BGo~film starts with the \Sone, \Stwo, and \SG~positions and widths derived from the XSW parameters. Fits starting with only \Sone~and \Stwo~always have an order of magnitude higher $\chi^2$ values and will not be discussed here.\cite{MConrad_thesis}  The higher $\chi^2$ values, using only two carbon layers in the buffer, supports our early argument that a 5-component fit to the C1s spectrum is necessary to get the correct vertical position of \Sone~and \Stwo~from the XSW analysis. The ML film uses a single component for the \BGML~film as determined from the preceding XSW analysis. Note that the XRR determined ML coverage is less than 1\% for the \BGo~film [see Fig.~\ref{F:Z-Profiles}(a)], consistent with the Raman ML estimate in Fig.~\ref{F:Raman}. 

Like the XSW profiles, the XRR analysis shows that the \BGo~layer distribution is very broad compared to the \BGML~layer in the ML film.  Associated with the change in the buffer's $z$-width, we find that the buffer-Si distance gets larger by 0.3\AA~ when a ML has grown above it.  This vertical change in distance results in a physically reasonable volume conservation when one also considers the previously observed transition from the larger in-plane incommensurate lattice spacing of \BGo~to the smaller commensurate lattice in the case of \BGML.\cite{Conrad_NL_17}  It apparently occurs concomitantly with a change in the distribution of buffer-Si bonds at the interface.  Both the change in width and bond distance are consistent with ARPES measurements that show a change in the band structure of the buffer when a ML grows above it.\cite{Conrad_NL_17} 

The XRR analysis finds that the bonding component \Sone~in the \BGo~layer is 1.9\AA~above the top Si layer; the same distance measured by STEM.\cite{Nair_NL_17}  The distance between the buffer \Sone~carbon and the last Si layer is essentially the same as the bulk Si-C bond (1.89\AA) confirming that the bond between \Sone~carbon and substrate Si has a significant covalent component.\cite{Emery_SW_13,Nair_NL_17}  We note that the XRR derived distance for \Sone~is about 17\% closer to the interface than the XSW result.  As discussed in Sec.~\ref{S:XSW} this small difference is not unusual given that the top Si-layer position is determined from the XSW yield of the Si 2s spectrum that has a significant contribution from deeper bulk-like bilayers with different vertical laxations.  Indeed, while the XSW analysis give the C-Si separation in the top SiC bilayer to be 0.69\AA, nearly the bulk value (0.63\AA), the XRR analysis finds that the C-Si separation has relaxed to be nearly 30\% shorter (0.47\AA) than in the bulk [see Table \ref{T:SXRD}].

The XRR derived vertical density profile of the \BGo~layer is both sharper and wider than the XSW distribution.  Figure \ref{F:Z_XrayvsSW} shows a comparison of the XRR and XSW derived distributions. The change in the sharpness of the \BGo~distribution is due to a $\sim\!60\%$ narrowing of the XRR derived \Sone~width compared to the XSW results [see Table \ref{T:SXRD}]). The XRR determined \BGo~layer width is $2.4(4)$\AA~compared to $1.5(2)$\AA~(including $\sigma$s) found in the XSW analysis.  We point out that both our XRR and XSW \BGo~widths are much wider than the previous XSW results of Emery et al.\cite{Emery_SW_13} (FWHM=0.9\AA). The difference is not unexpected because the prior XSW results were from multi-layer graphene films with a large areal coverage of \BGML~buffer and a much smaller coverage of \BGo~buffer. Based on our ML results, the early UHV films, which were composed of primarily ML graphene with little \BGo buffer,\cite{Emery_SW_13} would have measured a buffer width that was a weighted average of a majority  \BGML~film with width $0.4(1)$\AA~and a minority contribution from the broad \BGo~film with width $2.4(4)$\AA. Assuming a 80\% ML film, an average buffer width of 0.8\AA~would have been measured; close to the 0.9\AA~that was measured in Ref.~[\onlinecite{Emery_SW_13}].

While we find a wide \BGo~layer, it is not unprecedented. Large vertical buffer layer widths have also been suggested by scanning tunneling microscopy (STM) measurements.  Chen et al.,\cite{Chen_SS_2005} argue that large amplitude height variations in the \six buffer structure, seen in both filled and empty state images, are topographical.  They find vertical oscillations between 1.5\AA~and 3.2\AA. The larger widths were observed in films that were annealed for longer times, implying that they are associated with more ordered buffer films. However, similar oscillations seen by Riedl et al.,\cite{Riedl_PRB_07} were interpreted as being partially due to electronic effects so it remains difficult to make comparisons of our BGo~layer width with STM measured corrugations.  Ab initio calculations, using a bulk terminated SiC surface, predict a significant \BGo~layer modulations of 1.2\AA.\cite{Varchon_PRB_2008} As we now discuss, the interface is far from bulk terminated and could induce much larger theoretical modulations if a more realistic SiC interface structure was used. 
 \begin{figure}
\includegraphics[angle=0,width=7.0cm,clip]{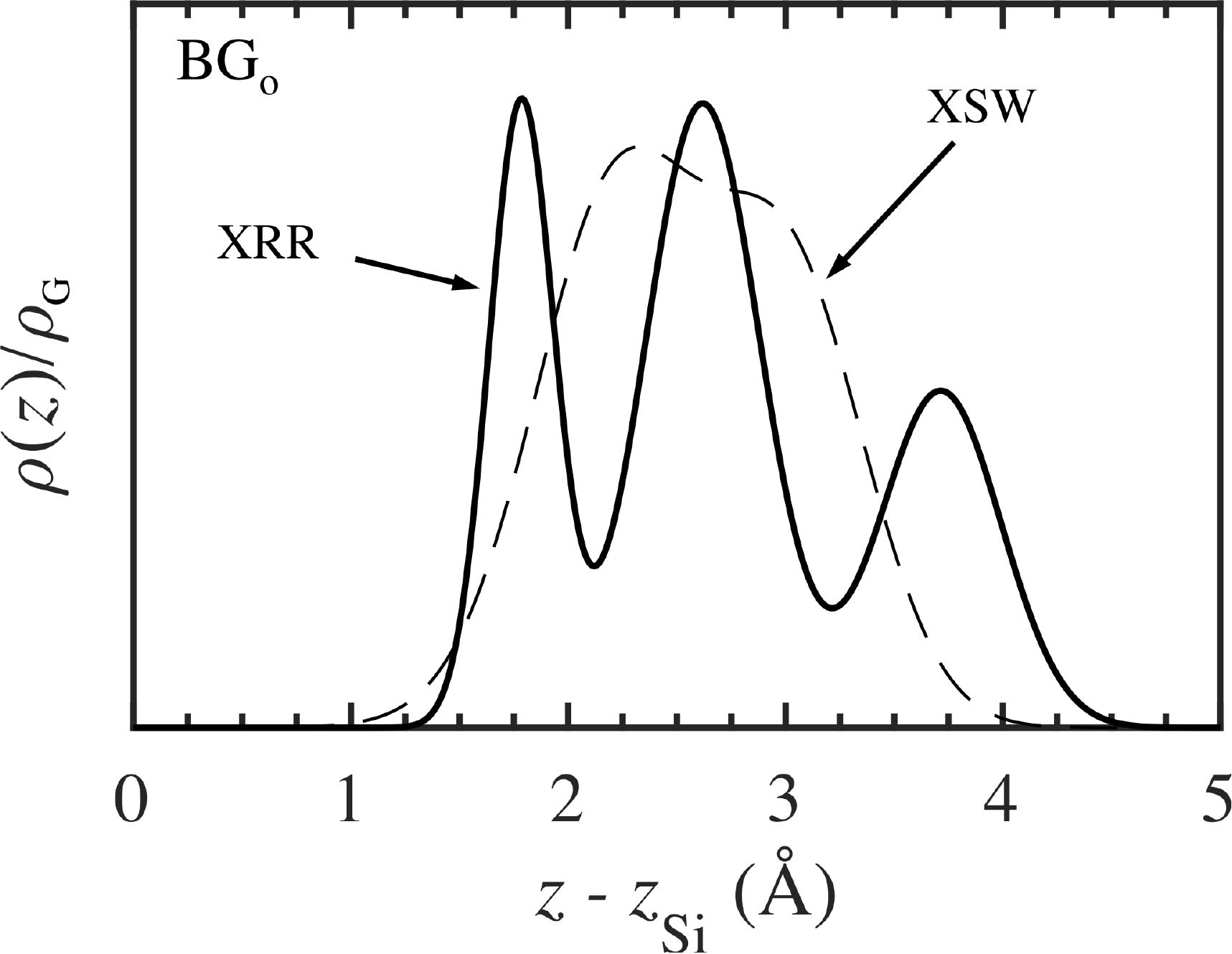}
\caption{A comparisons of the \BGo~layer's vertical density distribution, $\rho (z)$, derived from XSW (dashed line) and XRR (solid line). The density is normalized to the density of a graphene layer. }
\label{F:Z_XrayvsSW}
\end{figure}

The structure of the SiC interface layer and how Si in this layer is bonded to buffer carbon are the most important questions about this system. The reflectivity shows that while the last Si layer remains relatively sharp, its density is only 75\% of its bulk value (i.e.,  20\% Si vacancies) [see Table \ref{T:SXRD}].  A similar Si vacancy concentration was reported in previous XRR studies using UHV grown samples and was presumed to be due to a growth artifact.\cite{Emery_SW_13}  Our results strongly suggest that the depleted top Si layer is in fact an equilibrium structure.  We say this because the depleted fraction in the top Si layer is independent of the growth method. The Si vacancy concentration in our high temperature, high Si vapor pressure environment is the same as the concentration found in the prior low temperature, low Si vapor pressure UHV environment.\cite{Emery_SW_13}  Furthermore, the Si vacancy concentration is the same in both the \BGo~and \BGML+ML films [see Table \ref{T:SXRD}] even though the ML grows at a temperature 150\degC higher than the \BGo~layer.  
All of these results point to the depleted top Si layer being an equilibrium structure that is relatively independent of temperature and Si vapor pressure. The idea that Si vacancies may exist in the buffer-SiC system has been explored theoretically. Calculations have shown that Si vacancies or C interstitial in the interface layer below the \BGo~layer lead to lower total surface energies that can give rise to a number of possible large \BGo~corrugations.\cite{Kageshima_JJAP_11}  


  Besides the high Si vacancy concentration, there is a significant change in the carbon-silicon spacing in the top SiC bilayer.  The C-Si $z$-spacing, $\Delta_\text{Si-C}$, in the SiC bilayer below \BGo~is 30\% shorter compared to a bulk SiC bilayer and half the distance predicted by ab initio calculations in a bulk terminated surface with a \BGML+ML film.\cite{Sforzini_PRL_15}   It is likely that the vacancies in the top SiC bilayer lead to the additional bulk peak $\text{C}'_\text{B}$ in the 5-component C1s spectra in Figs.~\ref{F:XPScomp}(b) and (d). We suggest that not only are Si vacancies 
a part of the equilibrium SiC-buffer system but that they most likely drive the incommensurate buffer-SiC structure recently found in surface x-ray diffraction studies.\cite{Conrad_NL_17}  We expect that future theoretical work will support this assertion. 

We point out that while the XSW analysis finds that there is equal \Stwo~and \Sone~carbon [see Table \ref{T:SXRD}], the XRR derived coverages find that there are nearly twice as many \Stwo~carbon atoms compared to \Sone.  The higher \Stwo~coverage is more constant with the model in Fig.~\ref{F:Buffer_bonding} that would suggest approximately a 3:1 ratio of \Stwo~to \Sone.  Since the \Sone~bonding component, derived from the XRR, makes up 26\% of the carbon in the \BGo~layer [see Table \ref{T:SXRD}], the \Sone~coverage must put an upper limit on the number of Si-graphene bonds in the buffer layer.  If each \Sone~carbon atom bonds with a single Si atom in the top SiC interface layer, the reflectivity would estimate that 81\% of the interfacial Si is bonded to the buffer layer.
Of course it is not only the number of bonds that determine the \BGo 's electronic structure, but how these bonds are distributed.  Ab intio calculations of the \BGo~layer find that 25\% of the buffer carbon is bonded to 78\% of the Si in the top layer assuming a bulk terminated surface.\cite{Kim_PRL_08}  Tight binding calculations using an incommensurate distortion of the bulk surface predict a much lower number of bonded buffer carbon (15\%).\cite{Conrad_NL_17} It is very likely that allowing Si vacancies in the top SiC layer will lead to an opening of a gap in the band structure of buffer graphene in ab initio models.

Finally, a comparison of Figs.~\ref{F:Z-Profiles}(a) and (b) reveals that, with the exception of the last Si layer, the widths of the interface C and Si remain relatively well ordered after the buffer has formed ($\sigma_\text{C}=0.12$\AA~and $\sigma_\text{Si}=0.05$\AA).  Once the ML is grown at a higher temperature, there is considerably more vertical disorder in the interface. In the last two SiC bilayers, the C widths doubles compared to the buffer-only interface and the Si width in the second bilayer triples.  The increased vertical disorder is consistent with an increase in the in-plane disorder of the interface and an increase in the \BGML~in-plane strain when the ML grows.\cite{Conrad_NL_17}  Since the ML is grown 150\degC higher than the buffer only film, entropic disorder in the SiC below the ML film may become important.  This suggest that annealing studies (at temperatures less than the ML growth temperature) will need to be carried out to see if the interface order can be further improved.

\section{Conclusion}
In this work, we have studied the structure of the first graphene layer that grows on SiC(0001) (known as ``buffer" graphene).  By using samples with highly uniform and controlled graphene coverages, we are able to show that the buffer's structure is dramatically altered when monolayer graphene is grown above it. These results correlate well with electronic changes that occur with ML growth. From essentially ML free samples, we are able to clear up inconsistencies in early works as to which component in the buffer's C1s spectrum is associated with carbon bonded to Si in the SiC interface. We show that the \Sone~peak at a binding energy of 285.2~eV is from buffer carbon $\text{sp}^3$ bonded to Si. 


One of the most important findings of this work is that the SiC interface below the buffer graphene cannot be bulk terminated.  Instead we show that the last Si layer in the SiC interface is substantially reconstructed.  By comparing samples grown under different growth conditions, we show that the top Si bi-layer in the interface has a large equilibrium concentration of Si vacancies (20\%
of the bulk value). 
The effect of the reduced number of Si atoms is to decrease the top SiC bilayer distance by 30\% of the bulk value.  This planarization of the SiC bilayer may explain why the SiC interfacial layer, along with the buffer layer, becomes incommensurate with the bulk SiC. We also find that less than 26\% of the \BGo~buffer layer carbon is bonded to the substrate. While the exact density of Si in the top layer cannot be determined, we can report that the number of Si atoms bonded to the buffer carbon can be no more than  55\%
 of the Si atoms in the top SiC bilayer.  We believe these results will act as a new starting SiC interfacial structures for future ab initio calculations that will help understand the semiconducting properties of this graphene film. 

  We also show that the vertical corrugation of the buffer layer is very large, $\Delta Z\!=\!2.4(4)$\AA~and that the buffer's width reduces to $\Delta Z\!=\!0.4(1)$\AA~when a monolayer grows above the buffer.  The reduction in the buffer's width is associated with a slight increase in the Si-buffer carbon bond length.  Finally, we find that the SiC interface becomes more vertically disordered when the monolayer forms.  This correlates with previous x-ray diffraction results that find an increase in the buffer's in-plane disorder when a monolayer forms above it.\cite{Conrad_NL_17} This is most likely due entropic disorder at the higher growth temperature of the monolayer and the decrease of Si out-diffusion caused by the buffer carbon film.  The results suggest that annealing may improve the interface order.

 \vspace*{2ex}
\noindent\textbf{Acknowledgments}
This research was supported by the National Science Foundation under Grant No. DMR-1401193. P.F. Miceli also acknowledges support  from the NSF under grant No DGE-1069091.   Additional support came from the Partner University Fund from the French Embassy. The authors would like to thank W. de Heer for use of the Keck Lab facilities.  Finally, we wish to thank the staff and technical support given for this project by the Synchrotron Soleil under project No's 20150681 and 20150682.

\vspace*{2ex}

 \bibliography{refs}




\newpage

\end{document}